\documentclass[showpacs,twocolumn,aps,prb]{revtex4}
\usepackage{amsmath}
\usepackage{amsfonts}
\usepackage{amssymb}
\usepackage{amsthm}
\usepackage{graphicx}

\begin{document}

\title{Entanglement Spectrum and Entanglement Thermodynamics
of Quantum Hall Bilayers at $\nu=1$}
\author{John Schliemann}
\affiliation{Institute for Theoretical Physics, University of Regensburg,
D-93040 Regensburg, Germany}
\date{November 2010}

\begin{abstract}
We study the entanglement spectra of bilayer quantum Hall systems at total
filling factor $\nu=1$. In the interlayer-coherent phase at layer
separations smaller than a critical value, the entanglement
spectra show a striking similarity to the energy spectra of the
corresponding monolayer systems around half filling. 
The transition to the incoherent phase can be followed in terms of
low-lying entanglement levels. Finally, we describe the connection
between those two types of spectra in terms of an effective temperature
leading to relations for the entanglement entropy which are in full
analogy to canonical thermodynamics.
\end{abstract}
\pacs{73.43.-f,73.43.Nq}

\maketitle

\section{Introduction}

Experimental studies of quantum Hall bilayers at total filling factor 
$\nu=1$ have revealed a series of intriguing phenomena
\cite{Murphy94,Spielman00,Eisenstein,vKlitzing,Pellegrini,Atsugi}. From a
theoretical point of view, such strongly interacting systems pose 
intricate and subtle problems of many-body physics, and many unexpected
observations are still lacking a fundamental understanding \cite{MoonYang}. 
Here the exact numerical treatment of small systems has provided important
insights and guidance of theoretical intuition
\cite{He93,Schliemann01,Yoshioka,Shibata06,Nakajima02,Schliemann03,Simon03,Moller}.
Moreover, in the last decade  many-body physics has been enriched
by the concept of entanglement, whose extensive and systematic study originated
in the field of quantum information theory \cite{Amico08}. In particular,
the notion of the {\em entanglement spectrum} has led to novel insights
in the physics of quantum Hall monolayers at fractional filling factors
\cite{Li08,Lauchli10,Thomale10,Sterdyniak10}, spin systems of one
\cite{Calabrese08,Pollmann10a,Pollmann10b,Thomale09,Poilblanc10}
and two \cite{Yao10} spatial dimensions,
quantum hall insulators \cite{Fidkowski10,Prodan10}, and also
rotating Bose-Einstein condensates \cite{Liu10}.
The entanglement spectrum of a 
bipartite 
system is defined in terms of the Schmidt decomposition of its ground
state $|\psi\rangle$,
\begin{equation}
|\psi\rangle=\sum_{n}e^{-\xi_{n}/2}|\psi^{A}_{n}\rangle\otimes|\psi^{B}_{n}\rangle
\end{equation}
where the states $|\psi^{A}_{n}\rangle$ ( $|\psi^{B}_{n}\rangle$) form an
orthonormal basis of the subsystem A (B), and the nonnegative
quantities $\xi_{n}$ are the dimensionless levels of the 
entanglement spectrum. In the present paper
we study the entanglement spectrum of
quantum Hall bilayers at total filling factor $\nu=1$, where the subsystems
are naturally defined by the two layers. Our investigations are based on 
exact numerical diagonalizations in the spherical geometry \cite{Haldane83}.

\section{Results and discussion}

Depending on the layer separation and the amplitude for particle
tunneling between the layers, quantum Hall bilayers exhibit a 
quantum phase transition
between a phase showing spontaneous interlayer coherence and quantized
Hall conductance, and a phase where such effects are absent with the two layers
being merely uncorrelated. The corresponding phase diagram was first
mapped out experimentally by  Murphy {\it et al.} \cite{Murphy94} and later
reproduced theoretically via numerical diagonalizations \cite{Schliemann01}.
The latter study relied on a careful analysis of fluctuations of the
pseudospin describing the layer degree of freedom. Moreover, in the case
of of very small tunneling amplitude, the phase transition is
also signaled in experiment 
by the generation of a pronounced peak in the tunneling 
conductance at zero bias voltage \cite{Spielman00}. Here the critical value
of the layer separation of $d=1.83\ell$ ($\ell$: magnetic length) was nicely
reproduced by numerical calculations taking into account the finite width
of the two wells \cite{Schliemann01,Schliemann03,Burkov02}. 
In the following, however,
we shall for simplicity concentrate on the case of zero well width where the
transition is found numerically at $d=1.3\ell$, and we will focus
on the situation of vanishing tunneling amplitude. Moreover, we
will assume the physical electron spin (as opposed to its pseudospin) 
to be fully polarized by the perpendicular magnetic field such that
it does not play a role in the present study.

In the limit of zero layer separation $d=0$ both layers merge to a 
fully filled monolayer whose Coulomb ground state is given by the
well-known Slater determinant of the Vandermonde type,
\begin{equation}
|\psi\rangle\propto \left(\prod_{i<j}\left(z_{i}-z_{j}\right)\right)
\exp\left(-\frac{1}{4\ell^{2}}\sum_{i}|z_{i}|^{2}\right)|T^{z}\rangle\,,
\end{equation}
where the $z_{i}$ are the complex electron coordinates in the planar
geometry. An analogous wave function exists in the spherical model of
$N$ electrons where the sphere is penetrated by $N-1$ flux quanta
\cite{Haldane83}.
Here the $z$-component of the pseudospin can take values of
$T^{z}=-N/2,\dots N/2$ meaning that, say, the top/bottom layer
contains $N/2\pm T^{z}$ electrons. The case $T^{z}=0$ corresponds to
Halperin\rq s 111 wave function \cite{Halperin83}. Tracing out one of the
layers means summing over $N/2\pm T^{z}$ electron coordinates in the pure
state $\rho=|\psi\rangle\langle\psi|$ with $|\psi\rangle$ being a single Slater 
determinant such that the resulting $N/2$-body 
reduced density matrix is proportional to the unit matrix with all
eigenvalues being equal to $1/{N\choose N/2\pm T^{z}}$. Thus, at $d=0$ the
entanglement spectrum shrinks to a single value.

\begin{figure}
\begin{center}
\includegraphics[width=7.5cm]{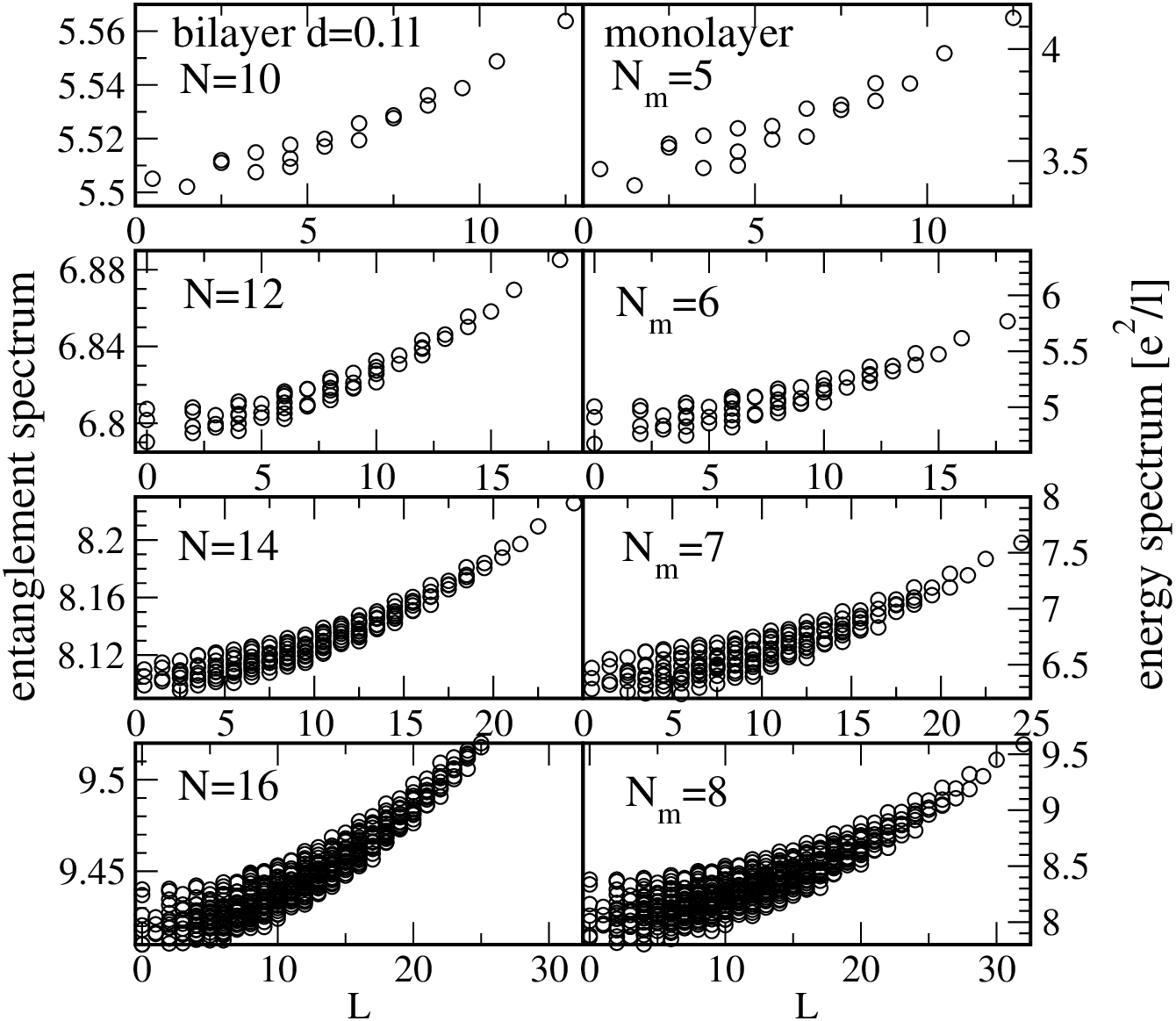}
\end{center}
\caption{Left: Entanglement spectra obtained from balanced bilayers 
at layer separation $d=0.1\ell$ for various system sizes $N$. Right:
Energy spectra of half-filled monolayers with $N/2$ electrons.
\label{fig1}}
\end{figure}

\begin{figure}
\begin{center}
\includegraphics[width=7.5cm]{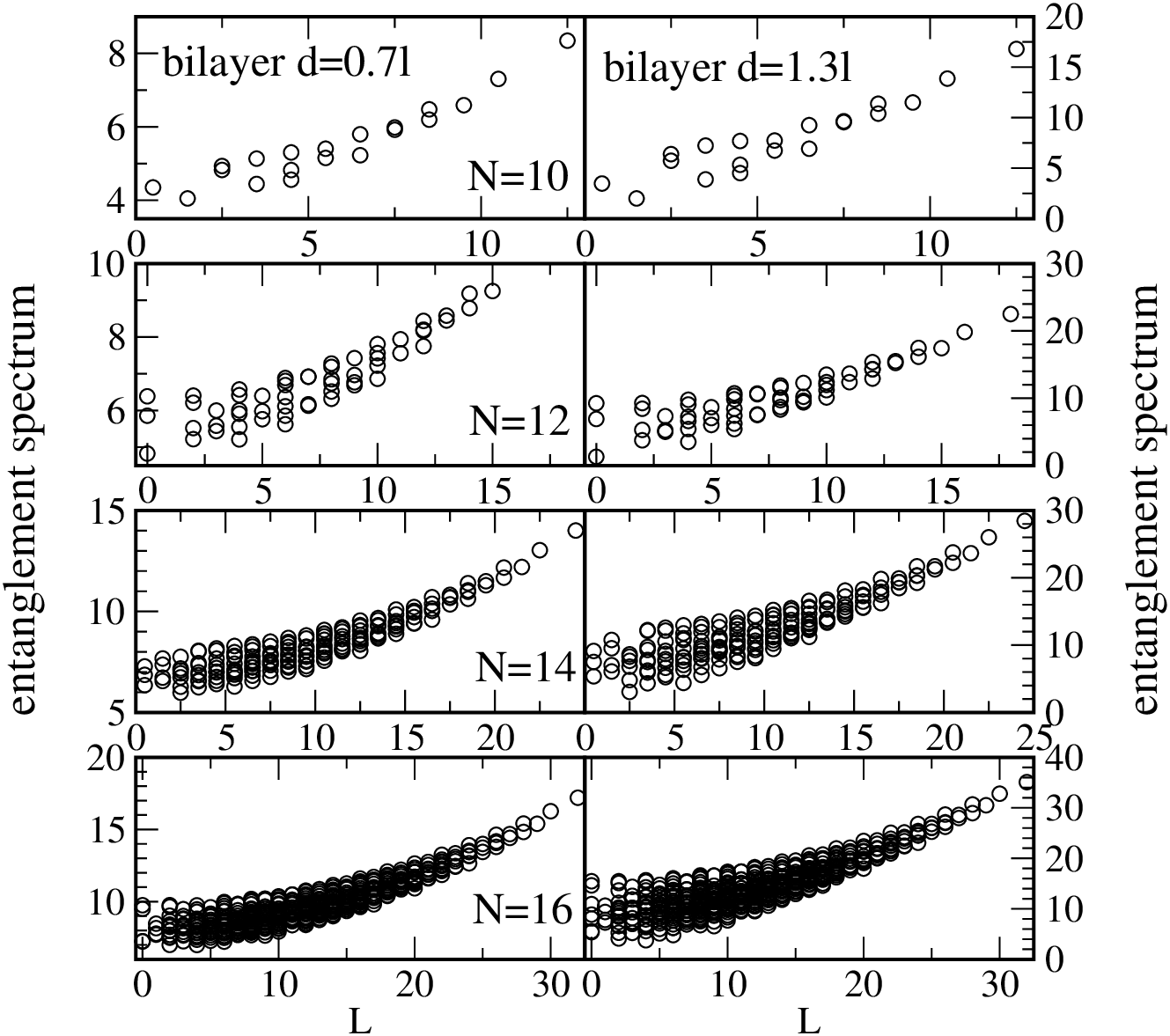}
\end{center}
\caption{Entanglement spectra of balanced bilayers 
at layer separation $d=0.7\ell$ (left) and $d=1.3\ell$ (right).
\label{fig2}}
\end{figure}

\begin{figure}
\begin{center}
\includegraphics[width=7.5cm]{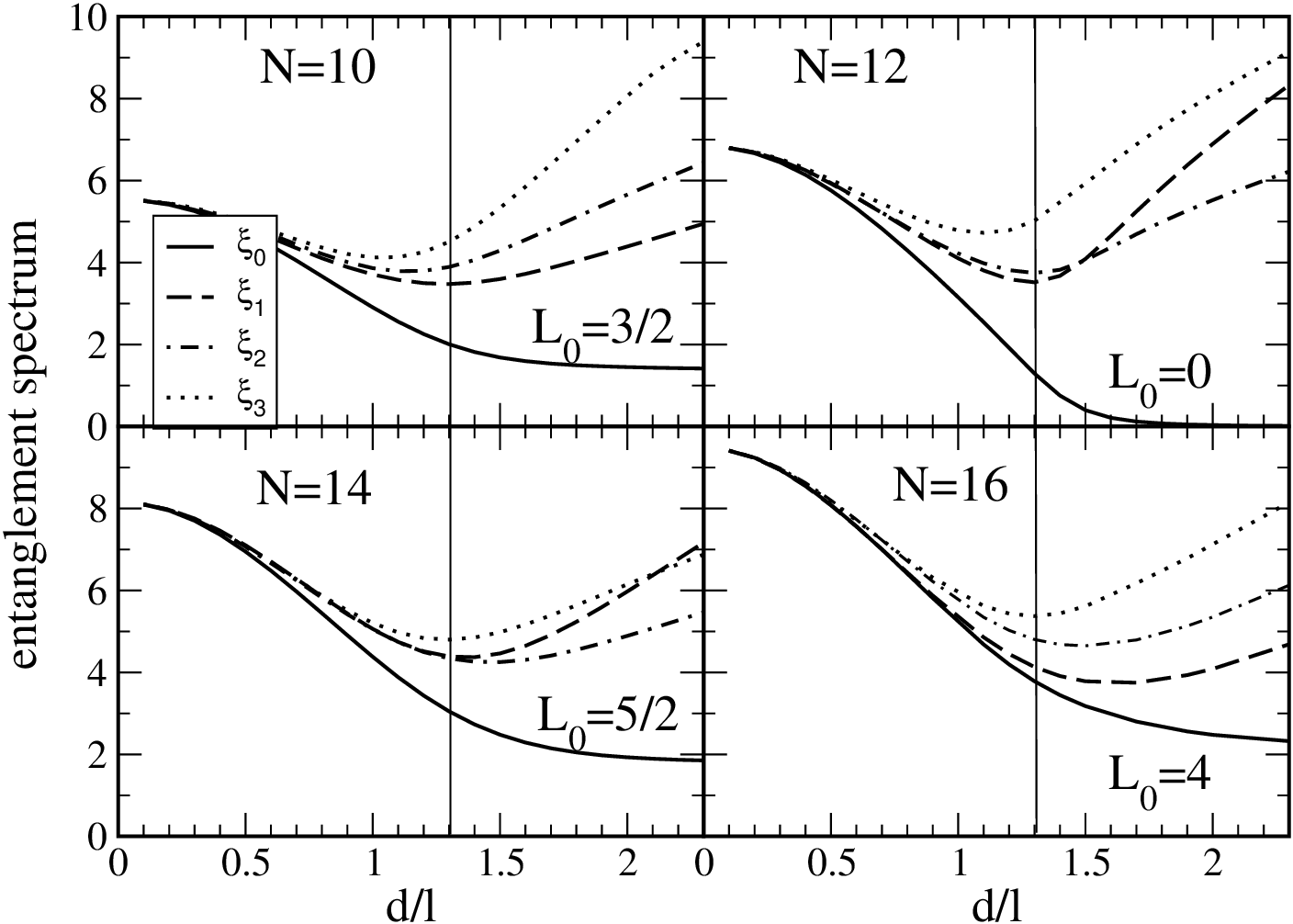}
\end{center}
\caption{The four lowest entanglement levels as a function of the
layer separation for different system sizes. At the phase boundary at
$d\approx 1.3\ell$ the entanglement spectrum qualitatively changes.
\label{fig3}}
\end{figure}

\begin{figure}
\begin{center}
\includegraphics[width=7.5cm]{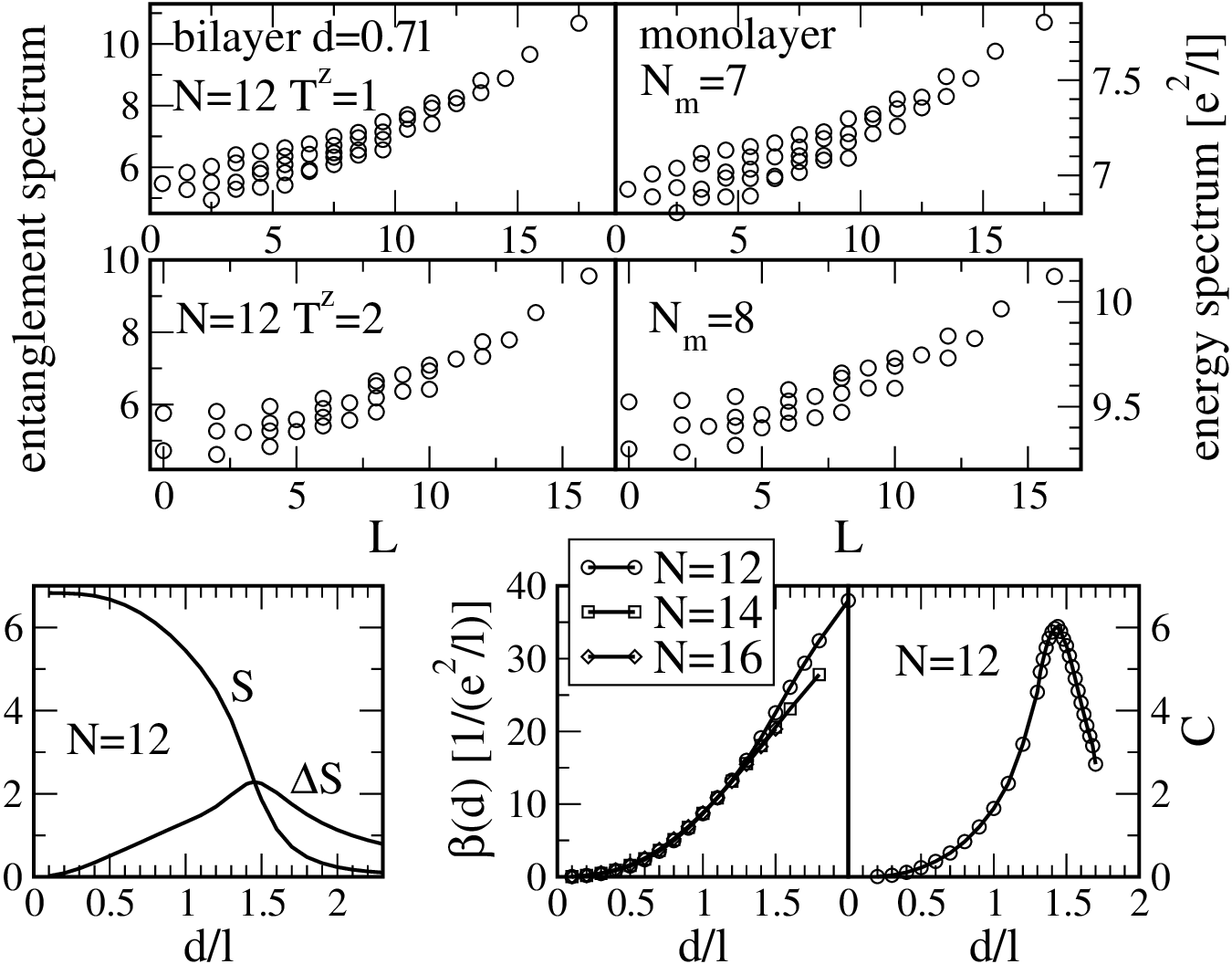}
\end{center}
\caption{Top panels: Entanglement spectra of 
$N=12$ electrons with $T^{z}\in\{1,2\}$ are compared with
the corresponding monolayer spectra .
Left bottom panel: Entanglement entropy $S$ along with its variance
$\Delta S$ as a function of the layer separation $d$ for a balanced bilayer 
system of $N=12$ electrons. Right bottom panels: The ``inverse temperature''
$\beta$ and the ``specific heat'' $C$ 
as a function of layer separation.  
\label{fig4}}
\end{figure}

Let us now turn to the case of finite $d>0$. The Coulomb ground state of
of the bilayer system in spherical geometry has vanishing total 
angular momentum
$L=0$. Therefore, the reduced density matrix is also rotationally invariant,
and its eigenvalues occur in multiplets of the angular momentum. Fig.~\ref{fig1}
presents entanglement spectra obtained from balanced bilayers ($T^{z}=0$)
at small but finite layer separation $d=0.1\ell$ for various
system sizes $N$ and compares them with the {\em energy spectra} of
half-filled monolayers containing $N_{m}=N/2$ particles. As seen from the figure,
both types of spectra show a striking similarity which is, on the total
scale of the spectra, particularly
evident at smaller system sizes.
This observation
also continues to larger layer separations as shown in Fig.~\ref{fig2}
for the cases $d=0.7\ell$  and $d=1.3\ell$: With increasing 
layer separation the entanglement spectra widen up (while shrinking to a 
single level for $d\to 0$), but remain similar in shape to the
monolayer energy spectra.
At layer separations
exceeding the critical value $d=1.3\ell$ a qualitatively different behavior
sets in: The global shape of the entanglement spectrum remains
reasonably close to the monolayer energy spectrum, but differences are
developed in details of the data.
This can be seen most clearly by following the lowest entanglement level
$\xi_{0}$ and the three following 
levels $\xi_{n}$, $n\in\{1,2,3\}$, as a function of the layer
separation\cite{note1}. 
Those entanglement levels (along with their counterparts in
the monolayer energy spectra) occur at different values of the angular
momentum $L_{n}$ which depend on system size and are summarized in table
\ref{table1}. 
\begin{table}
\begin{tabular}{c|c|c|c|c|}
N & $L_{0}$ & $L_{1}$ & $L_{2}$ & $L_{3}$ \\
10 & 3/2 & 1/2 & 7/2 & 9/2 \\
12 & 0 & 4 & 2 & 3 \\
14 & 5/2 & 11/2 & 7/2 & 5/2 \\
16 & 4 & 2 & 6 & 0 \\
\end{tabular}
\caption{Angular momentum $L_{n}$ of the low-lying entanglement levels
$\xi_{n}$ at various system sizes $N$ of the underlying bilayer system.
\label{table1}}
\end{table}
Fig.~\ref{fig3} shows these low-lying entanglement levels as a function of
the layer separation. While at $d\lesssim 1.3\ell$ all levels decrease
with increasing layer separation, the properties of the spectrum
change qualitatively at $d\approx 1.3\ell$: Here the excited levels start 
to increase with the layer separation while the lowest level
approaches $\xi_{0}\approx\ln(2L_{0}+1)$. Indeed, at large layer separations the
reduced density matrix of a single layer is dominated by a multiplet
with total angular momentum $L_{0}$ with all other eigenvalues being 
exponentially close to zero. In this sense, the ground state of the
bilayer system carrying angular momentum quantum numbers $L=L^{z}=0$ 
can at large layer separation be viewed to be composed of two 
monolayers with $L=L_{0}$ according to a standard Clebsch-Gordan
decomposition \cite{Landau,Schliemann00}, 
\begin{equation}
|0,0\rangle\approx\sum_{m=-L_{0}}^{L_{0}}\frac{(-1)^{L_{0}-m}}{\sqrt{2L_{0}+1}}
|L_{0},m\rangle|L_{0},-m\rangle\,,
\end{equation}
leading to a reduced density matrix of the form
\begin{equation}
\rho_{red}\approx\sum_{m=-L_{0}}^{L_{0}}\frac{1}{2L_{0}+1}
|L_{0},m\rangle\langle L_{0},m|\,.
\end{equation}
Thus, the entanglement entropy $S=\langle-\ln\rho_{red}\rangle$,
$\langle\cdot\rangle:={\rm tr}(\cdot\rho_{red})$,
interpolates, similarly as the lowest entanglement level
$\xi_{0}$, between a value of $S=\ln{N\choose N/2}$ at $d=0$ and
$S=\ln(2L_{0}+1)$ at large layer separation. In the left bottom panel of
Fig.~\ref{fig4} we have plotted $S$ along with its variance
\begin{equation}
\Delta S=\sqrt{\langle(-\ln\rho_{red}-S)^{2}\rangle}
\end{equation}
as a function of $d$
for a bilayer system of $N=12$ electrons.
The entropy $S$ shows an inflection point, accompanied by a maximum of 
$\Delta S$, near $d=1.4\ell$, which is close to the finite-size value
of the phase boundary obtained earlier from an analysis of the 
pseudospin fluctuations \cite{Schliemann01,note2}.

The observation that the entanglement spectrum of 
interlayer-coherent quantum Hall  bilayers 
shows an intriguing similarity to the the energy spectrum of the
corresponding monolayer extends also to unbalanced systems. This is illustrated
in the top panels of Fig.~\ref{fig4} where the entanglement spectra of 
$N=12$ electrons with $T^{z}\in\{1,2\}$ obtained by tracing out the top
layer are compared with
the corresponding monolayer spectra. 
Entanglement spectra
obtained from bilayer systems with negative $T^{z}$
(or, alternatively, by tracing out the bottom layer)
are related to the previous ones
by a particle-hole transformation are therefore identical, and the 
corresponding monolayer energy spectra just differ
by an additive constant.

The investigations so far have concentrated on the case of
vanishing single-particle tunneling between the layers. At finite
tunneling amplitude, the particle number of each layer is no longer a 
good quantum number, and the monolayer Fock space consists of invariant
subspaces of the reduced density matrix which are characterized both by total
angular momentum and particle number. Therefore, for not too large
tunneling, the entanglement spectra within each subspace of given
particle number will be similar to those given in 
Figs~\ref{fig1},\ref{fig2},\ref{fig3} up to an additive constant
describing the spectral weight of the respective subspace.

Focusing again on the case of zero tunneling, our results 
suggest that the reduced density matrix fulfills, for not too large
layer separation $d$, the following approximate relation,
\begin{equation}
-\ln\rho_{red}\approx\beta{\cal H}_{m}+\ln Z\,,
\label{fit}
\end{equation}
where $Z={\rm tr}\exp(-\beta{\cal H}_{m})$, and 
${\cal H}_{m}$ describes the Coulomb repulsion in a
half-filled monolayer with $N$ flux quanta and $N_{m}=N/2$ electrons. 
The (inverse) ``entanglement temperature'' $\beta$ is a
parameter depending on $d$, and the occurrence of the partition function
$Z$ in Eq.~(\ref{fit}) ensures the condition ${\rm tr}\rho_{red}=1$.
Clearly, $d\to 0$ implies $\beta\to 0$ and $Z={N\choose N/2}$.
The mid bottom panel of Fig.~\ref{fig4} shows
least-mean-square results for $\beta$ as a function of
layer separation. As seen there, the data depend only very weakly on system 
size. Moreover, defining a ``entanglement free energy'' by
$F=-T\ln Z$, $T=1/ \beta$, one derives from Eq.~(\ref{fit}) the familiar
thermodynamic relation
\begin{equation}
F\approx E-TS
\label{frerg}
\end{equation}
with $E=\langle{\cal H}_{m}\rangle$.
Thus, the entanglement entropy of the bilayer system turns out to have
an immediate thermodynamic meaning for the monolayer system
characterized by the parameter $T=1/ \beta$.
These findings suggest to define, in full analogy with
canonical thermodynamics, a ``specific heat'' via
\begin{equation}
C=T\frac{dS}{dT}=-\beta\frac{dS}{d\beta}\,.
\label{spech}
\end{equation}
In the mid bottom panel of Fig.~\ref{fig4} we have plotted
$C$  as a function of layer separation for a
system of $N=12$ electrons. as seen, the specific heat
has a pronounced peak near the phase boundary which should be seen as a
strong hint for the underlying phase transition being first order.
Finally, from Eq.~(\ref{fit}) one also derives the result
\begin{equation}
\left(\Delta S\right)^{2}\approx\beta^{2}\left(
\langle{\cal H}_{m}^{2}\rangle-
\langle{\cal H}_{m}\rangle^{2}\right)
\end{equation}
which provides a direct link between the variance of the entanglement
entropy of the bilayer system and the energy variance of the monolayer.
As seen in Fig.~\ref{fig4}, $\Delta S$ shows a maximum at the phase boundary
which should be considered as another piece of evidence 
that, in the infinite-volume limit, the
phase transition is of first order \cite{Schliemann01} rather than 
continuous \cite{Shibata06}, i.e. the entropy changes discontinuously.

Very recently, Poilblanc has performed a study of entanglement spectra
of Heisenberg spin ladders, where a similarly striking resemblance
to the energy spectra of single spin chains was found \cite{Poilblanc10}.
This observation could also be described in terms of an effective
temperature. 
The results of both papers raise the
question of a common underlying mechanism being at work, also extending
previous studies on fractionally filled quantum Hall monolayers
\cite{Li08,Lauchli10,Thomale10,Sterdyniak10} to bilayers at filling 
factor $\nu=1$ investigated here. Most interestingly, the latter direction
of work makes
immediate contact to the enigmatic system of a half-filled
quantum Hall monolayer. However, here we should add the caveat that
the monolayer systems that naturally arise in our present work contain
$N$ electrons in $N_{\phi}=2N-1$ flux quanta whereas 
in many previous numerical studies on
(electron-spin polarized) monolayers 
at half filling considered using spherical
geometry other relations of the form $N_{\phi}=2N-M$, $M$ being a small 
integer, have been considered \cite{1/2}.
To eliminate such a finite-size
``shift problem'' it would be desirable to repeat
the calculations reported on here in the toroidal geometry.

We note that a different recipe of how to eliminate finite-size properties
from entanglement spectra of quantum Hall monolayers in the spherical geometry
was put forward very recently by Thomale {\it et al.} \cite{Thomale10}. 
These authors propose to strip all normalization factors from single-particle
states entering the Slater determinants of many-body wave functions. This
procedure of achieving a ``conformal limit'' conserves azimuthal symmetry
on the sphere (as appropriate for the situation of Ref.~\cite{Thomale10}),
but destroys full rotational covariance, which is desired 
in the present study on bilayer systems.
Thus, the prescription proposed in Ref.~\cite{Thomale10} is 
unfortunately not applicable to our investigations here.

\section{Conclusions}

In summary, we have studied the entanglement spectra of bilayer quantum Hall 
systems at total filling factor $\nu=1$. Our investigations are based on
exact numerical diagonalizations using the spherical geometry.
In the interlayer-coherent phase at small layer separations, the entanglement
spectra show a striking similarity to the energy spectra of the
corresponding monolayer systems around half filling. 
The transition to the incoherent phase can be followed in terms of
low-lying entanglement levels, constituting a link between the entanglement
spectrum and a quantum phase transition. Clear signatures of the quantum
phase transition are also shown by the entanglement entropy along with
its fluctuation. Moreover, the connection
between those two types of spectra can be described in terms of an 
effective temperature which gives rise to
relations for the entanglement entropy being fully analogous to
canonical thermodynamics. In particular,
the specific heat derived from this formalism provides a strong hint
for the phase transition being of first order \cite{Schliemann01}.

\acknowledgments{This work was supported by Deutsche Forschungsgemeinschaft 
via SFB 631.}



\begin{thebibliography}{}

\bibitem{Murphy94}
S.~Q. Murphy {\it et al.}, Phys. Rev. Lett. {\bf 72}, 728 (1994).

\bibitem{Spielman00}
I.~B. Spielman {\it et al.}, Phys. Rev. Lett. {\bf 84}, 5808 (2000).

\bibitem{Eisenstein}
M. Kellogg {\it et al.}, Phys. Rev. Lett. {\bf 88}, 126804 (2003);
{\it ibid.} {\bf 90}, 246801 (2003);
{\it ibid.} {\bf 93}, 036801 (2004); 
A.~R. Champagne {\it et al.}, {\it ibid.}  {\bf 100}, 096801 (2008);
A.~D.~K. Finck {\it et al.}, {\it ibid.} {\bf 104}, 016801 (2010).

\bibitem{vKlitzing}
Y. Yoon {\it et al.}, Phys. Rev. Lett. {\bf 104}, 116802 (2010);
R.~D. Wiersma {\it et al.}, {\it ibid.} {\bf 93}, 266805 (2004).

\bibitem{Pellegrini}
S. Luin {\it et al.},  Phys. Rev. Lett. {\bf 90}, 236802 (2003);
B. Karmakar {\it et al.}, {\it ibid.} {\bf 102}, 036802 (2009);
arXiv:0907.4032.

\bibitem{Atsugi}
N. Kumada, {\it et al.}, Phys. Rev. Lett. {\bf 94}, 096802 (2005);
P. Giudici {\it et al.}, {\it ibid.} {\bf 100}, 106803 (2008);
{\it ibid.} {\bf 104}, 056802 (2010).

\bibitem{MoonYang}
K. Moon {\it et al.}, Phys. Rev. B {\bf 51}, 5138 (1995); 
K. Yang {\it et al.}, {\it ibid.} {\bf 54}, 11644 (1996).

\bibitem{He93}
S. He, S. Das Sarma, and X.~C. Xie, Phys. Rev. B {\bf 47}, 4394 (1993).

\bibitem{Schliemann01}
J. Schliemann, S.~M. Girvin, and A.~H. MacDonald,
Phys. Rev. Lett. {\bf 86}, 1849 (2001).

\bibitem{Yoshioka}
K. Nomura and D. Yoshioka,  Phys. Rev. B {\bf 66}, 153310 (2002);
N. Shibata, Prog. Theor. Phys. Supp. {\bf 176}, 182 (2008). 

\bibitem{Shibata06}
N. Shibata and D. Yoshioka, J. Phys. Soc. Jpn. {\bf 75}, 043712 (2006).

\bibitem{Nakajima02}
T. Nakajima, Phys. Rev. B {\bf 65}, 233317 (2002).

\bibitem{Schliemann03}
J. Schliemann, Phys. Rev. B {\bf 67}, 035328 (2003).

\bibitem{Simon03}
S.~H. Simon, E.~H. Rezayi, and M.~V. Milovanovic,
Phys. Rev. Lett. {\bf 91}, 046803 (2003).

\bibitem{Moller}
G. M\"oller, S.~H. Simon, and E.~H. Rezayi, 
Phys. Rev. Lett. {\bf 101}, 176803 (2008);
Phys. Rev. B {\bf 79}, 125106 (2009).

\bibitem{Amico08}
L. Amico {\it et al.}, Rev. Mod. Phys. {\bf 80}, 517 (2008).

\bibitem{Li08}
H. Li and F.~D.~M. Haldane, Phys. Rev. Lett. {\bf 101}, 010504 (2008).

\bibitem{Lauchli10}
A.~M. L\"auchli {\it et al.}, Phys. Rev. Lett. {\bf 104}, 156404 (2010).

\bibitem{Thomale10}
R. Thomale {\it et al.},  Phys. Rev. Lett. {\bf 104}, 180502 (2010).

\bibitem{Sterdyniak10}
A. Sterdyniak, N. Regnault, and B.~A. Bernevig, arXiv:1006.5435.

\bibitem{Calabrese08}
P. Calabrese and A. Lefevre, Phys. Rev. A {\bf 78}, 032329 (2008).

\bibitem{Pollmann10a}
F. Pollmann and J.~E. Moore, New J. Phys. {\bf 12}, 025006 (2010).

\bibitem{Pollmann10b}
F. Pollmann {\it et al.}, Phys. Rev. B {\bf 81}, 064439 (2010).

\bibitem{Thomale09}
R. Thomale, D.~P. Arovas, and B.~A. Bernevig,
Phys. Rev. Lett. {\bf 105}, 116805 (2010).

\bibitem{Poilblanc10}
D. Poilblanc, Phys. Rev. Lett. {\bf 105}, 077202 (2010).

\bibitem{Yao10}
H. Yao and X.-L. Qi, Phys. Rev. Lett. {\bf 105}, 080501 (2010).

\bibitem{Fidkowski10}
L. Fidkowski, Phys. Rev. Lett. {\bf 104}, 130502 (2010).

\bibitem{Prodan10}
E. Prodan, T.~L. Hughes, and B.~A. Bernevig, 
Phys. Rev. Lett {\bf 105}, 115501 (2010).

\bibitem {Liu10}
Z. Liu {\it et al.}, arXiv:1007.0840.

\bibitem{Haldane83}
F.~D.~M. Haldane, Phys. Rev. Lett. {\bf 51}, 605 (1983).

\bibitem{Burkov02}
A. Burkov {\it et al.}, Physica E {\bf 12}, 28 (2002).

\bibitem{Halperin83}
B.~I. Halperin, Helv. Phys. Acta {\bf 56}, 75 (1983).

\bibitem{note1}
We do not show here complete entanglement spectra for layer separations
larger than $d=1.3\ell$ since in such situations, in particular at larger
system sizes, the high-lying entanglement levels correspond to
exceedingly small eigenvalues of the reduced density matrix whose
precise computation tends to overdemand the available numerical
accuracy. The low-lying entanglement levels shown in Fig.~\ref{fig3} are
not affected by this problem.

\bibitem{Landau}
See, e.g., L.~D. Landau and E.~M. Lifshitz, {\em Quantum Mechanics},
Butterworth Heinemann 2005.

\bibitem{Schliemann00}
A similar scheme of adding angular momenta of single layers was also
found for the electronic spin in quantum Hall bilayers at total
filing factor $\nu=2$,
J. Schliemann and A.~H. MacDonald, Phys. Rev. Lett. {\bf 84}, 4437 (2000).

\bibitem{note2}
We expect those slightly differing finite-size values for the phase boundary
to converge in the thermodynamic limit.

\bibitem{1/2}
See, e.g., Refs.~\cite{Moller} and 
E. Rezayi and N. Read, Phys. Rev. Lett. {\bf 72}, 900 (1994);
R.~H. Morf, {\it ibid.} {\bf 80}, 1505 (1998).

\end{thebibliography}
\end{document}